\documentclass[twocolumn]{aastex631}
\usepackage{natbib}
\usepackage{comment}
\usepackage{hyperref}
\usepackage{lineno}

\graphicspath{{./}{figures/}}
\usepackage{makecell}
\usepackage{booktabs}
\usepackage{ulem}

\usepackage{placeins}
\usepackage{xcolor}
\definecolor{ForestGreen}{RGB}{34, 139, 34}

\definecolor{Violet}{RGB}{148, 0, 211} 

\definecolor{MyRed}{RGB}{220, 20, 60} 

\shorttitle{Classifying mergers with deep learning}
\shortauthors{Lee et al.}

\begin{document}

\title{Convolutional Neural Networks for classifying galaxy mergers: Can faint tidal features aid in classifying mergers? }

\author[0009-0008-4978-9053]{Yeonkyung Lee}
\affil{Department of Astronomy, Space Science and Geology, Chungnam National University, 99 Daehak-ro, Yuseong-gu, Daejeon 34134, Republic of Korea}

\author[0000-0002-4362-4070]{Hyunmi Song}
\affil{Department of Astronomy, Space Science and Geology, Chungnam National University, 99 Daehak-ro, Yuseong-gu, Daejeon 34134, Republic of Korea}

\author[0000-0001-5135-1693]{Jihye Shin}
\affiliation{University of Science and Technology (UST), Gajeong-ro, Daejeon 34113, Republic of Korea}

\author[0000-0001-9991-8222]{Sungryong Hong}
\affiliation{Korea Astronomy and Space Science Institute, 776 Daedeokdae-ro, Yuseong-gu, Daejeon 34055, Republic of Korea}

\author[0000-0002-6810-1778]{Jaehyun Lee}
\affiliation{Korea Astronomy and Space Science Institute, 776 Daedeokdae-ro, Yuseong-gu, Daejeon 34055, Republic of Korea}

\author[0000-0001-9544-7021]{Kyungwon Chun}
\affiliation{Korea Astronomy and Space Science Institute, 776 Daedeokdae-ro, Yuseong-gu, Daejeon 34055, Republic of Korea}

\correspondingauthor{Hyunmi Song}
\email{lee8yklee@gmail.com, hmsong@cnu.ac.kr}

\begin{abstract}
Identifying mergers from observational data has been a crucial aspect of studying galaxy evolution and formation.
Tidal features, typically fainter than 26 ${\rm mag\,arcsec^{-2}}$, exhibit a diverse range of appearances depending on the merger characteristics and are expected to be investigated in greater detail with the Rubin Observatory Large Synoptic Survey Telescope (LSST), which will reveal the low surface brightness universe with unprecedented precision. Our goal is to assess the feasibility of developing a convolutional neural network (CNN) that can distinguish between mergers and non-mergers based on LSST-like deep images.
To this end, we used Illustris TNG50, one of the highest-resolution cosmological hydrodynamic simulations to date, allowing us to generate LSST-like mock images with a depth $\sim$ 29 ${\rm mag\,arcsec^{-2}}$ for low-redshift ($z=0.16$) galaxies, with labeling based on their merger status as ground truth.
We focused on 151 Milky Way-like galaxies in field environments, comprising 81 non-mergers and 70 mergers.
After applying data augmentation and hyperparameter tuning, a CNN model was developed with an accuracy of 65--67\%.
Through additional image processing, the model was further optimized, achieving an accuracy of 67--70\% when trained on images containing only faint features.
This represents an improvement of $\sim$ 5\% compared to training on images with bright features only.
This suggests that faint tidal features can serve as effective indicators for distinguishing between mergers and non-mergers.
The future direction for further improvement based on this study is also discussed.
\end{abstract}

    
\section{Introduction} \label{sec:intro}

Hierarchical mergers of small structures play a key role in the formation of cosmic structures in the $\Lambda$CDM universe \citep[e.g.,][]{White1978, White1991, Kauffmann1993, GuoandWhite2008, Conselice2014}. As directly observable probes, galaxy mergers provide key insights into the formation and evolution of cosmic structures and cosmological models.
The astrophysical implications of mergers are also significant, considering that mergers cause rapid and significant changes to galaxies \citep[e.g.,][]{Dubois2016,Martin2018,MartinJ2020,Davison2020,Remus2022,Cannarozzo2023}. 
Simulations that trace the merger process over time have shown that both star formation activity and accretion onto supermassive black holes (SMBHs) are often enhanced shortly before and after mergers \citep[e.g.,][]{Springel2016,Thorp2019,Montero2019}.
These activities appear to diminish as gas is rapidly depleted or ejected.
Additionally, the merger process can greatly alter the gas distribution in and around galaxies through gas inflow and feedback from stars or active galactic nuclei \citep[e.g.,][]{Satyapal2014,Goulding2018,Ellison2019,Byrne-Mamahit2023}.
To confirm these theoretical/numerical predictions, it is essential to establish tools that can identify galaxies involved in mergers and infer information about their merger stages based on observational data.

One way to identify galaxy mergers from observational data is by locating galaxy pairs that are spatially close to each other \citep[e.g.,][]{Barton2000,Lin2004}.
Galaxies that are close in both celestial coordinates and redshift space are likely to interact or merge in the near future.
However, this approach is suitable for identifying pre-merger systems.
To find galaxies in the post-merger stage,  many observations have been made to identify galaxies characterized by multiple cores, asymmetric morphology, and tidal features.

Tidal features are a key indicator of on-going or post mergers, but their quantification is challenging due to their variety of forms \citep[e.g.,][]{Johnston1999,Johnston2008,kawata2006,Mancillas2019,Khaldi}.
Therefore, visual inspection is still widely used to identify these features and determine whether a merger has occurred.
However, visual inspection is not easily applicable for vast datasets. 
While projects like Galaxy Zoo \citep{Lintott2011} managed to involve the public in visually inspecting Sloan Digital Sky Survey (SDSS) data \citep{York2000}, this approach suffers from the subjectivity and variability of individual judgements, as well as being time-consuming. 
Given that upcoming surveys like the Rubin Observatory Legacy Survey of Space and Time \citep[LSST;][]{LSST2019} will produce data on an even larger scale than SDSS, visual inspection alone is unlikely to be a viable method for such massive datasets.

For better efficiency and consistency, research has begun applying machine learning algorithms like convolutional neural networks (CNNs) to image-based classification tasks \citep[e.g.,][]{Ackermann2017,Jacobs2019,HuertasCompany2018,HuertasCompany2019,HuertasCompany2020,Bottrell2019,Pearson2019,Walmsley2019,ReimanGohre2019,Cheng2020,Ferreira2020,Ferreira2022,MartinK2020,Walmsley2020,Wang2020,Bickley2021,Bottrell2022,Ferreira2024,Bickley2024b,Chudy2025,deGraaff2025}.
CNNs extract information from galaxy images through convolutions with various filters, optimizing hyperparameters to ensure that the extracted features are highly correlated with pre-defined labels.
As a result, CNNs can automatically extract features strongly related to the label without explicitly parametrizing morphological characteristics.
In this process, a key factor is labeling since the model's performance and obejctive heavily depend on it.

For example, \citet{Ackermann2017} and \citet{Pearson2019} built a CNN model using SDSS data that had been labeled as merger or non-merger through visual inspection.
The resulting model achieved an exceptionally high accuracy of over 90\%. 
However, mergers identified by this model may be biased toward those that leave visually distinct morphological distortions, as the labeling is based on visual inspection. 
Therefore, this model may be more appropriate for detecting tidal features or morphological distortions rather than for identifying mergers.
In this sense, simulation data that provides the ground truth of merger history is more suitable for building a CNN model that can classifying bona fide mergers.
To identify galaxies involved in mergers from SDSS and JWST images, \citet{Pearson2019} and \citet{Ciprijanovic2020} built a CNN model based on simulation data.
The accuracy of these models ranged from 65\% to 87\%, which is lower than models trained using labels from visual inspection.
This suggests that the classification of mergers and non-mergers based solely on morphological features is challenging.
The decreasing accuracy is partly due to the impact of flybys which can distort the morphology of neighboring galaxies during close encounters \citep[e.g,][]{Prodanovi2013,Kim2014,Lang2014}.

These previous studies have developed CNN models for classifying mergers based on morphological features visible at a depth limit of around 25 ${\rm mag\,arcsec^{-2}}$. 
For high-redshift galaxies, this depth limit restricts the visible features primarily to the galaxy's central regions. 
Even for low-redshift galaxies, tidal features in the outer regions are not always clearly visible, and given the pixel scale of SDSS, these features are not fully resolved in detail. 
Compared to SDSS \citep{miskolczi2011}, LSST will have twice the pixel resolution ($0.2\,\mathrm{arcsec\,pixel^{-1}}$) and is expected to reach a surface brightness limit that is four magnitudes deeper (with a 3$\sigma$ surface brightness limit of $\sim29$ ${\rm mag\,arcsec^{-2}}$) in its 10-year average observations \citep{Seppo2018}.
This will allow LSST to unveil the low surface brightness universe in unprecedented detail, revealing both prominent and subtle signs of various interactions between galaxies.
However, it is not immediately clear whether this abundance of information will aid in merger classification, as galaxy interactions that do not involve mergers can also produce merge-like features.

Several studies have highlighted LSST's capability in detecting tidal features and identifying mergers using simulation data.
\citet{Martin2022}, using the New Horizon simulation \citep{Dubois2021}, showed that LSST will detect $\sim$60--80\% of tidal features in Milky Way-like galaxies at $z\sim0.05$ and that these features will remain observable up to intermediate redshifts ($z<0.2$).
\citet{Bickley2024b}, using TNG100 \citep{Springel2018}, demonstrated that higher image quality improves merger identification, with LSST outperforming surveys such as SDSS, DECaLS, CFIS, and HSC-SSP.
Although matching LSST's surface brightness limit requires simulation data with exceptionally high resolution, only a few studies have taken advantage of TNG50 \citep{Nelson2019}, one of the most advanced cosmological hydrodynamic simulations to date.
This is largely due to its relatively small volume, which limits the number of available galaxies--a potential drawback for machine learning models that require large training sample.
Nevertheless, TNG50's superior resolution makes it uniquely suited for capturing faint tidal features, which can be crucial for identifying galaxy mergers in deep images surveys.

In this study, we utilize TNG50 to train a CNN model for merger classification, with a goal of assessing whether the subtle tidal features it resolves--often under-represented in lower-resolution simulations--can aid in classifying mergers.
To maximize the model's accuracy, various hyperparameters of the CNN and image processing techniques (e.g., emphasizing or removing specific features) are optimized.
As a feasibility study, this study focuses on Milky Way-like central galaxies in field environments, but planning to widen the ranges of masses and environments as well as to include satellite galaxies in future work.

The remainder of the paper is structured as follows. 
In Section \ref{sec:Data}, we present the simulation data and the construction of mock images.
Section \ref{sec:method} describes the architectures of the CNN models both the fiducial model and improved models.
We then present and discuss the results of model training, highlighting the influence of faint tidal features on merger classification in Section \ref{sec:res}.
Finally, we conclude with a summary and an outlook for future work in Section \ref{sec:summary}.

\section{Data} \label{sec:Data}
\subsection{The galaxy sample and merger classification}
To develop a merger identifying CNN model, we utilized the Illustris TNG50 simluation \citep{Pillepich2019}, the highest-resolution model in the IllustrisTNG series and one of the most advanced cosmological hydrodynamic simulations to date.
The high resolution of TNG50 is critical to create mock images with surface brightness limit comparable to those of the LSST.

We used the $z=0.2$ snapshot data, with which we can identify on-going or future mergers by examining the merger tree at $z<0.2$.
We specifically focused on Milky Way-like galaxies, which are central galaxies in the range of $8\times10^{11} \leq M_{\rm halo}/M_\odot \leq 2\times10^{12}$. 
This choice was partly motivated by the goal of understanding the merger history of our own galaxy.
Additionally, by narrowing the sample, we aimed to reduce the complexity of the merger classification problem, making it more manageable, as a pilot study.
As a result, our sample includes 151 galaxies.

The mass assembly history of TNG50 galaxies can be traced using their merger trees, constructed using the tree-building algorithm \textsc{sublink} \citep{RodriguezGomez2015} or \textsc{LHaloTree} \citep{Springel2005}.
Although the two algorithms define the first (main) progenitors slightly differently, the most massive history \citep[for more details, see][]{DeLucia2007} in the case of \textsc{sublink} and the most massive halo for \textsc{LHaloTree}, they generally produce similar results.
To identify mergers in our target galaxies, we utilize the merger history catalog constructed by \citet{RodriguezGomez2017} and \citet{Eisert2023} based on \textsc{sublink}.

In the catalog, major mergers are defined as those with a stellar mass ratio greater than 1/4 and minor mergers as those with a mass ratio between 1/10 and 1/4.
These mergers are identified for various time windows, ranging from 250 Myr to 8 Gyr into the past, relative to a given epoch.
Given that tidal features can be produced by minor mergers \citep[e.g.,][]{DOnghia2009} and may persist for up to 3 Gyr \citep{Khaldi}, we defined mergers as galaxies that have undergone a merger with a stellar mass ratio greater than 1/10 (encompassing both the major and minor mergers, as defined in the catalog) within the last 2 Gyr.
Since a merger is identified at the moment when two progenitors coalesce, interacting progenitors that have not yet merged would be classified as non-mergers.
However, it is more appropriate to identify these systems as mergers (or more precisely, ongoing mergers). 
Therefore, we examined the future merger tree relative to the chosen snapshot (i.e., $z=0.2$) to account for these cases.
With these criteria, our target galaxies are classified into 81 non-mergers and 70 mergers (54 post-mergers and 16 ongoing mergers).
Although the sample size is not large enough, this limitation is partly resolved through data augmentation as described in the next section.

\begin{figure*}[ht!]
\centering
\includegraphics[width=\linewidth]{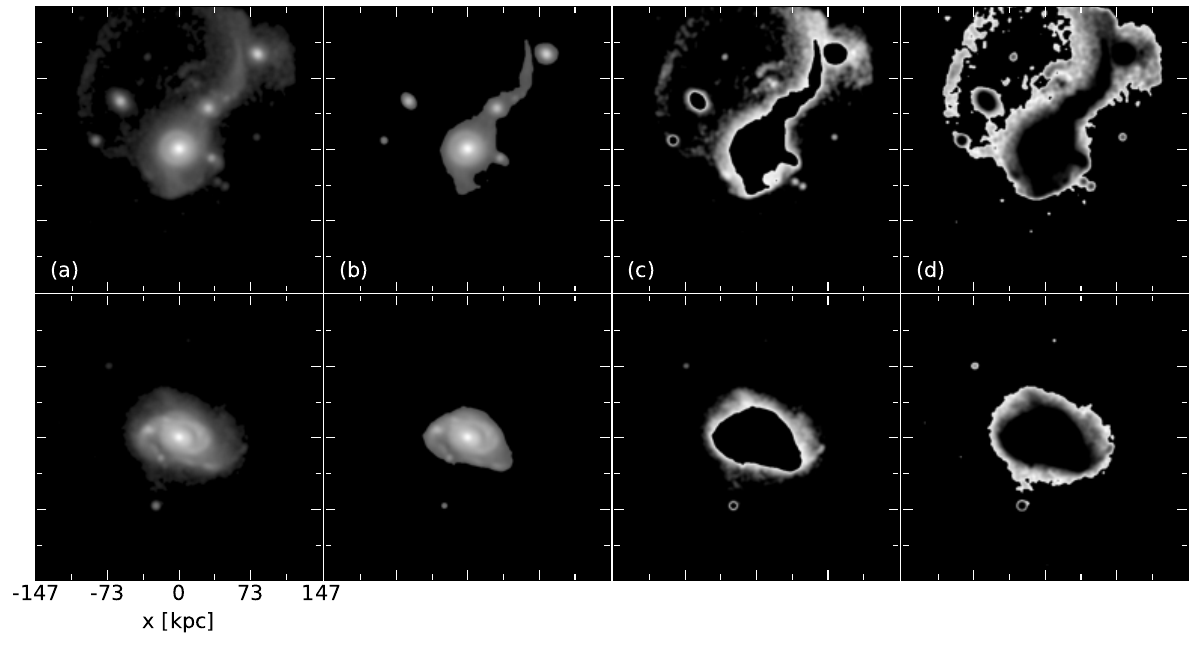}
\caption{The top and bottom rows display example galaxies with and without tidal features, respectively. Each column represents different image processing methods: (a) original images with no mask (NM), (b) images after masking faint features (MF), (c) images after masking bright features (MB), and (d) images after masking bright features and inverting unmasked, star-particle pixels (MBI).}
\label{fig:SBM}
\end{figure*}

\subsection{Surface Brightness Map}\label{sec:SBM}
To create surface brightness maps of our target galaxies, we included stellar particles within 20 effective radius ($20R_e$) from the galaxy center, aiming to fully capture the diffuse features in their outskirts.
We used the (rest-frame) $K$-band luminosity calculated for each star particle \citep{Trcka2022}, without the need to account for dust attenuation as $K$-band is less affected by dust extinction.
In contrast, dust attenuation must be considered at shorter wavelengths, a factor we plan to address in future work that will incorporate surface brightness maps across different wavelengths.
An additional benefit of using the $K$-band is that it effectively traces the overall stellar distribution, including tidal features. 
For this reason, even though $K$-band is not part of the LSST filters, we choose to use it for this feasibility study.

Considering the LSST pixel scale ($0.2\arcsec$), the 10-year surface brightness limit ($\sim29\,{\rm mag\,arcsec^{-2}}$ averaged across all bands) and the baryonic mass resolution of TNG50 ($8.5\times10^{4}M_\odot$), we determined the optimal distance to be $z=0.16$.
At this distance, pixels containing a single stellar particle reach the 10-year surface brightness limit.
While mimicking the effect of seeing ($\sim0.7\arcsec$, the fiducial value for the LSST survey), artifacts caused by the limited number of stellar particles, particularly the overestimates in the surface brightness of pixels with a single stellar particle, can be largely mitigated.
Although we selected galaxies from the snapshot at $z=0.2$ to track their evolution forward for up to 2 Gyr, we place them at $z=0.16$ when generating images.
This is the lowest redshift at which star particle pixels, on average, match the LSST surface brightness limit.
In this case, a pixel subtends 490.6 physical pc at $z=0.16$.

The surface brightness of each pixel is calculated following \citet[][see Eqs. (1)-(6)]{Tang2018}.
The surface brightness maps are sized at 600 by 600 pixels, corresponding to 294.4 physical kpc on each side.
We then convolved each map with a 2D Gaussian kernel corresponding to the fiducial seeing value.
We first processed images without background noise first, followed by those with background noise. 
We note that neighbouring and background galaxies beyond $20R_e$ from each target galaxy are not included in the maps.

We additionally processed the maps in three different ways.
The first approach applied a brighter surface brightness limit similar to the SDSS \citep{miskolczi2011}, which excludes faint features.
The second approach did the opposite, excluding bright features of $<26{\rm mag/arcsec^2}$.
Lastly, building on the second approach, we further modified the maps by inverting them, assigning higher values to fainter features when normalizing the maps between zero and one for input into a CNN model.
Figure  \ref{fig:SBM} shows example surface brightness maps processed using the three different approaches in addition to the original map.
By comparing the performance of CNN models trained on each of these maps, we can gain insight into which features, whether bright, faint, or inverted, are more relevant to galaxy mergers.

To overcome the limitations of the small sample size and projection effect, we augmented the mock images of each target galaxy by generating views from different projection angles.
One set undergoes mild augmentation with three projections along the $x$, $y$, and $z$ axes, while the other undergoes more aggressive augmentation, utilizing 28 different orientations determined by \textsc{HEALPix} \citep[]{healpix1, healpix2} with \texttt{nside}=1 (12 directions), which are then doubled by applying a 90-degree rotation.
A total of 453 and 4228 mock images are generated to develop a CNN model for merger classification.

We note that, as in other studies, each image is treated as an independent case, and images of the same galaxy may appear in the training, validation, and test sets, potentially introducing an overfitting issue.
Due to the small sample size, it is not feasible to fully separate galaxies across the training, validation, and test sets.
Nevertheless, as shown in the subsequent sections, the final model's performance remains stable across 1000 different realizations of the dataset splits, suggesting that overfitting may not be a significant concern.
Furthermore, when the sample size is increased by relaxing the mass range, thereby reducing the redundancy of a single galaxy across the splits, the model performance remains largely unchanged (see Section \ref{sec:additionaltests}).
Further studies with larger datasets and stricter isolation between subsets should validate these findings.

\section{Method: A Convolutional Neural Network for merger classification}\label{sec:method}
\subsection{The fiducial model}\label{sec:method:fiducial}

A CNN \citep{LeCun1998} is a machine learning algorithm specialized for image classification and feature detection.
CNNs extract key features through convolution layers and pooling operations, which can be used to classify galaxy mergers.
We use Gradient Class Activation Mapping \citep[Grad-CAMs;][]{grad-cam} to estimate relatively important regions for classification in images processed by CNNs, which allows us to visualize the relative importance of pixels of an image for the given task.
It can help interpret and understand the results of CNNs.

As our fiducial model, we adopted the CNN architecture from \citet{Ciprijanovic2020}, which was developed to identify mergers at high redshifts.
This provided a good starting point, as their objective closely aligns with ours--identifying mergers--but at different redshifts.
Since we are targeting faint tidal features in the outer region of galaxies at low redshifts, the image size needed to capture all relevant features is necessarily larger (in terms of the number of pixels) than that used in \citet{Ciprijanovic2020}.
To address this, we adjusted the stride and kernel sizes in the first convolutional layer, ensuring the input image size for the second layer matches that of \citet{Ciprijanovic2020}.
The model architecture is summarized in Table \ref{tab:Fiducial}.

The CNN model is trained for up to 500 epochs, with early stopping implemented to prevent overfitting.
Training is halted when no improvement in model performance is observed after the validation loss reaches its minimum.
The best model is chosen based on the highest validation accuracy achieved during training.

Building on this fiducial model we fine-tuned various hyperparameters (e.g., batch size, number of convolutional layers, and the splitting ratios for training, validation and test sets) and explored alternative options for the activation function, optimizer, and dilation to enhance model performance.
These adjustments are detailed in the following section.

\begin{figure*}[ht!]
\centering
\includegraphics[width=\linewidth]{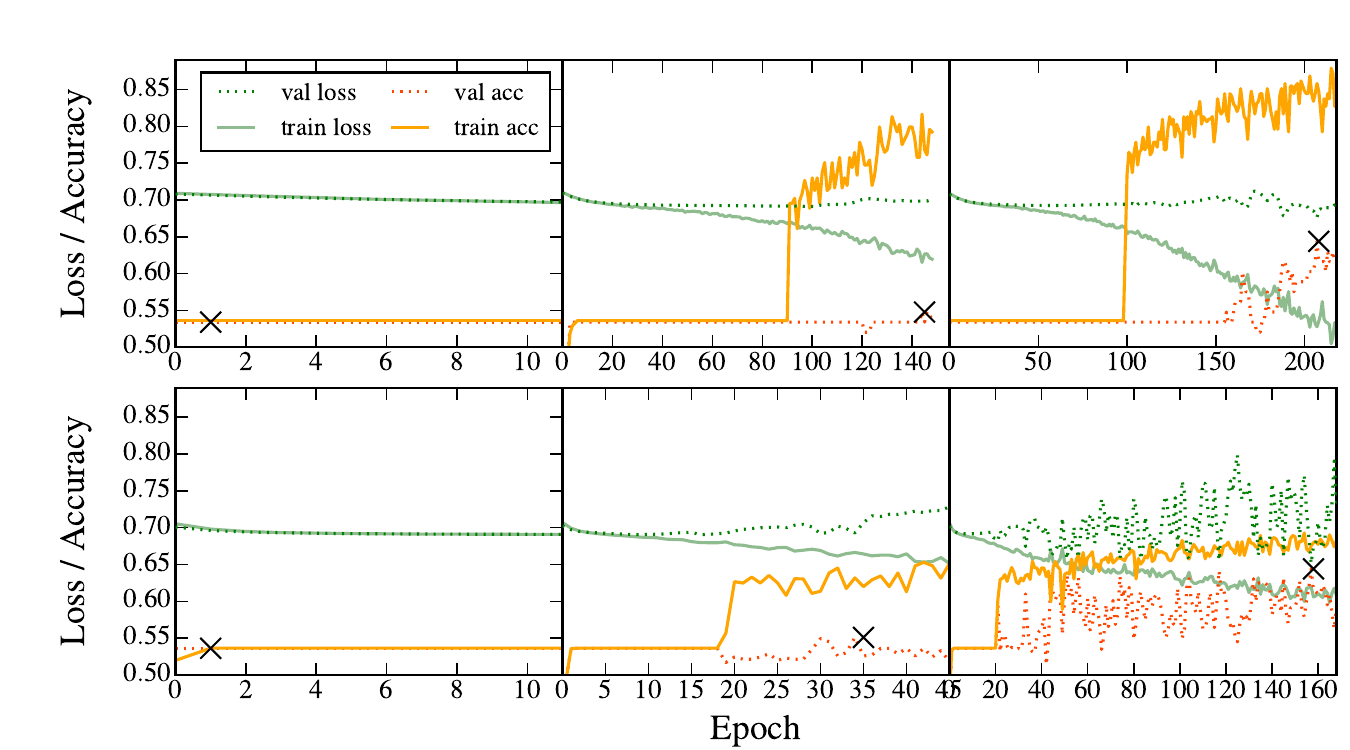}
\caption{Training curves for accuracy and loss of \texttt{Fiducial3} (top) and \texttt{Fiducial28} (bottom). Crosses represent the epoch with the highest accuracy, at which point the model's weights were saved for the final model.
Each column is representative case of the types of history curve: no training conducted (left), poor performance and/or lack of improvement on the validation set (middle), and effective training (right).}
\label{fig:history}
\end{figure*}

\subsection{Model improvements}\label{sec:method:hyperparams}
Hyperparameters are externally configured parameters that are manually set prior to training.
As they have a significant impact on model performance, it is crucial to find the optimal combination for a given dataset through extensive experiments.
In our experiments, we tested variations in activation function, dilation, batch size, optimizer, the number of convolutional layer, and the splitting ratios for training, validation and test sets.
The detailed architectures for these models are presented in Appendix \ref{app:model}.

An activation function determines how the output is transformed based on its input, playing a crucial role in capturing the non-linear relationships between inputs and outputs.
It applies a mathematical operation to the output that each neuron gives, introducing non-linearity into the model.
In the fiducial model, the Rectified Linear Unit (ReLU) is used, which converts negative values to zero.
While ReLU is one of the most popular activation functions, it can lead to issues such as the ``dying ReLu'' problem, where certain weights and biases of neurons are not updated.
This issue is mitigated by a variation called Leaky ReLU, which allows a small but non-zero gradient for negative inputs.
We considered ReLU and LeakyReLU when optimizing our model architecture.

An optimizer is a mathematical algorithm that adjusts the weights and biases of the network, enabling the efficient and stable minimization of the loss function.
We considered using the Adaptive Momentum Estimation (Adam) optimizer for the fiducial model, while Rectified Adam (RAdam) was tested as an alternative.
RAdam is particularly beneficial in preventing the model from falling into a local minimum due to large variations in the adaptive learning rate.
Batch size, the number of training subsets utilized in each iteration of the model training process, is also tested along with each optimizer.
The optimal batch size can vary across different optimizers, so we tested sizes of 64, 128 and 256.
Another critical hyperparameter to tune for an optimizer is the learning rate, which determines the size of the steps taken towards the minimum of the loss function.
While RAdam is less sensitive to changes in the learning rate \citep{Liu2020}, further testing is needed for Adam.
Therefore, we additionally tested learning rates of 0.01, 0.001 and 0.0005 for Adam.

We also adopted a dilated convolution that samples pixel values with spacing specified by a parameter called the dilation rate.
This approach increases the receptive field, enabling more comprehensive extraction of features, which is particularly useful to deal with large images.
We used a dilated convolution in the first layer and three dilation rates of 5, 10, and 15 were tested.
For each dilation rate, the stride and the kernel sizes were adjusted accordingly.

The number of convolutional layers, initially set to three in the fiducial model, was increased to four. 
We investigated whether the additional layer revealed any new merger-related features.
Layers beyond four were not considered, as this could lead to overfitting, particularly given the limited dataset.
The splitting ratios of training, validation and test sets were set at 64\%, 16\% and 20\% for the fiducial model, and adjusted to 48\%, 12\% and 40\% as an alternative.
While the latter combination may help prevent overfitting and provide a more stable evaluation due to the larger test set fraction, it requires caution as it may result in inadequate training.

By changing the abovementioned hyperparameters individually, we were able to understand which ones most significantly impact the model performance in merger classification.
Ultimately, we determined the optimal combination of the hyperparmaeters for constructing an enhanced CNN model.
Before fine-tuning the hyperparameters, we explored how data augmentation could improve the model using 453 and 4228 mock images described in Section \ref{sec:SBM}.
The performances of the models will be compared in Section \ref{sec:res:hyppar}, where the best model with the optimized configuration will also be presented.

\section{Results and discussion}\label{sec:res}
Each CNN model was evaluated using metrics such as training history, accuracy, F1-score, and Area Under Curve (AUC).
The training history shows the evolution of accuracy and loss for both the training and validation sets throughout the training process.
Based on the training history, we assessed the success of the training process and identified the final model at the epoch with the highest validation accuracy.
For each hyperparameter setting, 1000 models were trained using 1000 bootstrap-resampled datasets.
The performance of a given hyperparameter setting was determined by the median and standard deviation of the evaluation scores across the 1000 models.
While accuracy simply represents the fraction of the correct predictions, the F1-score, defined as the harmonic mean of precision and recall, offers a more reliable evaluation when class sizes are imbalanced and/or when the model performance varies significantly across different classes.
AUC measures the area under the Receiver Operating Characteristic (ROC) curve, which visualizes the relationship between the True Positive Rate (TPR) and False Positive Rate (FPR) for varying prediction thresholds.
A good model will have a low FPR and a high TPR, yielding an AUC value close to one.
Since a random model has an AUC of 0.5, a well-performing model should have a higher AUC than that.
AUC provides a comprehensive evaluation by assessing model performance across all possible thresholds, whereas accuracy and F1-score can be affected by the choice of a specific threshold.
because these three evaluation metrics are complementary to each other, we use all of them to assess model performance.

\subsection{Model optimization: hyperparameter tuning}\label{sec:res:hyppar}

Before presenting the results from hyperparameter tuning, we first demonstrate the model evaluation for fiducial models trained on the datasets augmented in two ways: mild augmentation, where images were multiplied by a factor of 3 using 3 different viewing angles, and aggressive augmentation, which involved multiplying by a factor of 28, as described in Section \ref{sec:SBM}.
The models are referred to as \texttt{Fiducial3} and \texttt{Fiducial28}, respectively.

Figure \ref{fig:history} presents example training curves for three distinct cases with \texttt{Fiducial3} ones shown in the top row and \texttt{Fiducial28} ones in the bottom row: where no training has been conducted (left), where the model demonstrates low performance and/or lack improvement on the validation set (middle), and where training has proceeded effectively (right).
The first two cases indicate unsuccessful model training: the first case is failed training, while the second case indicates overfitting, where the model has been trained but only performs well on the training dataset, failing to generalize to new data.
Among 1000 models, the distributions of these cases for \texttt{Fiducial3} are 64.8\%, 18\%, and 17.2\%, respectively.
For \texttt{Fiducial28}, the distributions are 7\%, 58\%, and 35\%.
It is worth noting that aggressive augmentation 
significantly reduces the fraction of the completely failed cases (from 64.8\% to 7\%).

\begin{figure*}[ht!]
\centering
\includegraphics[width=\linewidth]{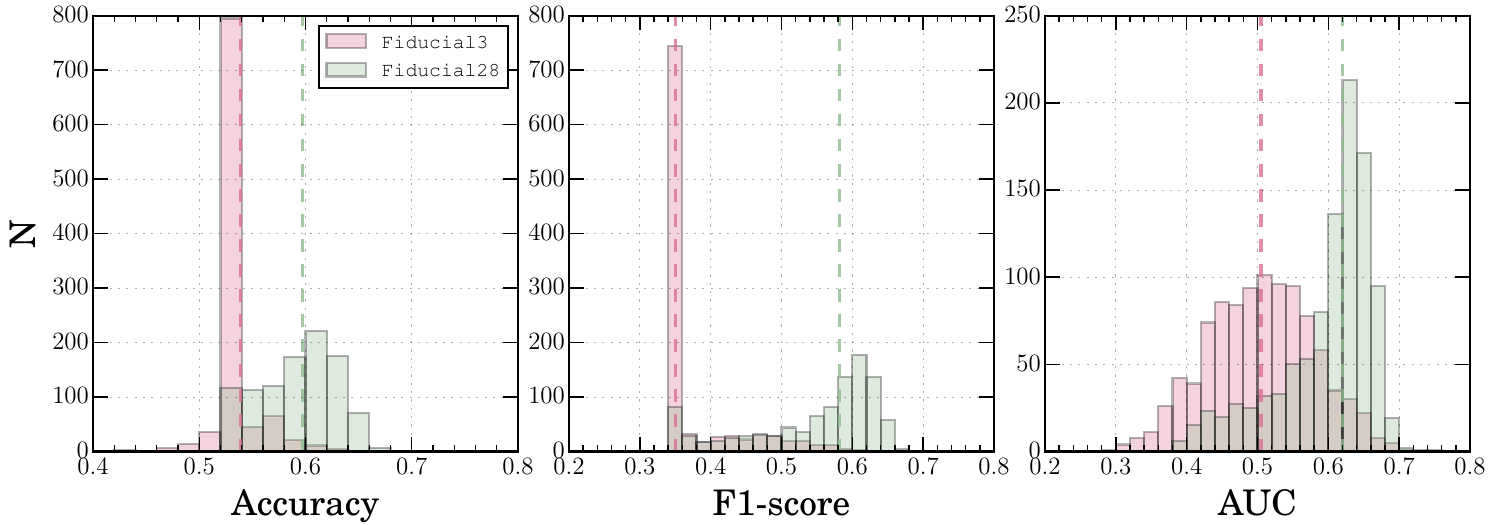}
\caption{Accuracy, F1-score, and AUC distributions of 1000 model instances of \texttt{Fiducial3} and \texttt{Fiducial28}. Pink and green dashed lines represent the median of each distribution.}
\label{fig:acchist}
\end{figure*}

The overall performance of \texttt{Fiducial3} and \texttt{Fiducial28} is summarized in Figure \ref{fig:acchist} and Table \ref{tab:accall}, which show the distributions of accuracy, F1-score and AUC for each set of 1000 model instances for each.
While the performance of \texttt{Fiducial3} appears comparable to that of a random classifier (i.e., its median AUC is $\sim0.5$), \texttt{Fiducial28} demonstrates a clear improvement.
These results suggest that the primary reason for the low performance of \texttt{Fiducial3} models is likely the small size of the dataset.
While the models of \texttt{Fiducial28} exhibit better performance than those of \texttt{Fiducial3}, the unsuccessful fraction (i.e., 7\%+58\%=65\%) remains significant.
Additionally, the overall performance is below the accuracy expected for a machine learning model (e.g., a minimum accuracy requirement of $\sim 60\%$), and appears unstable, exhibiting large variability.
It seems that there are limitations to model improvement through data augmentation, but there may be some potential for further enhancement through hyperparameter tuning.

\begin{figure*}[ht!]
\centering
\includegraphics[width=\linewidth]{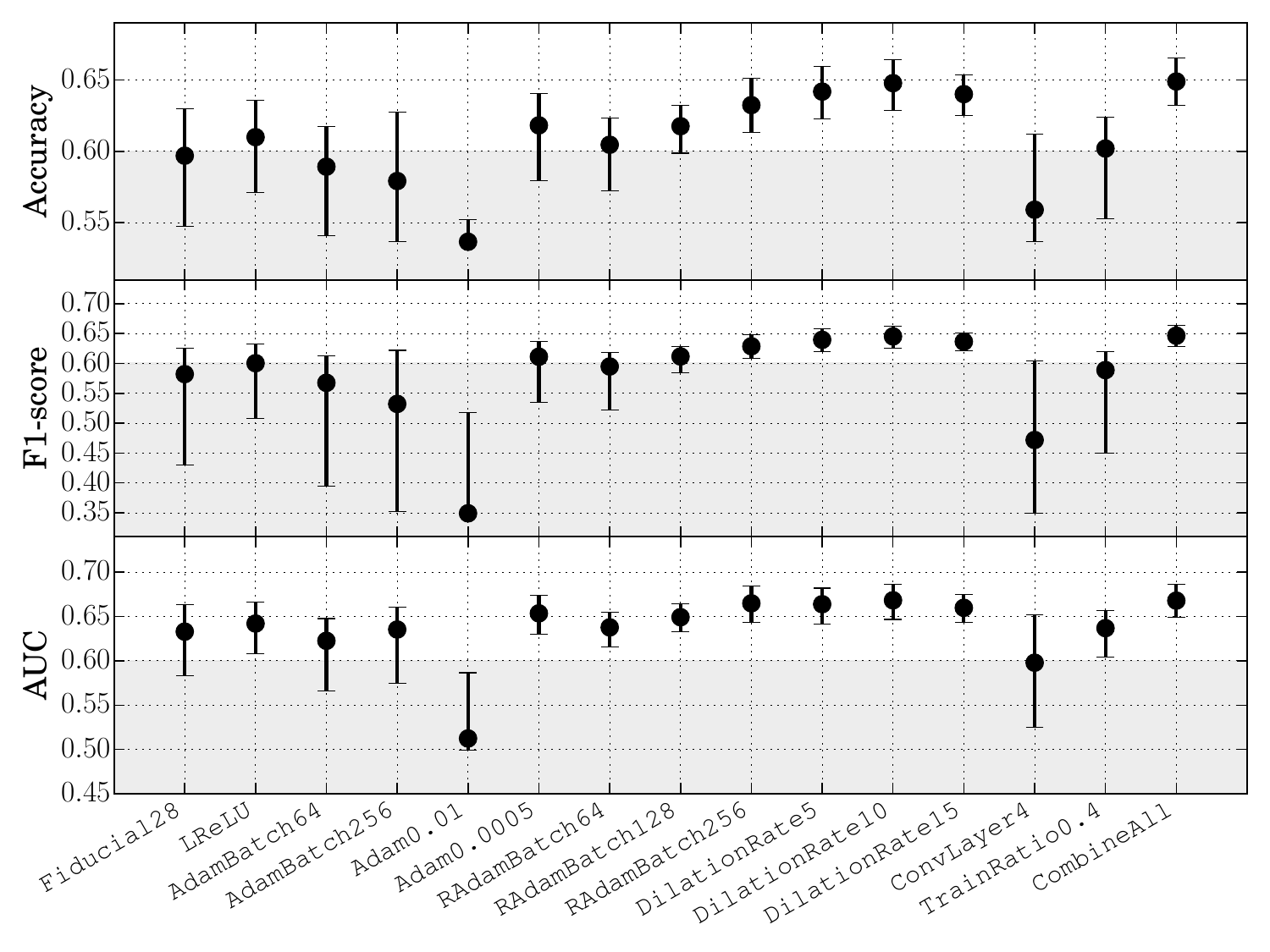} 
\caption{Comparison of model performance across different hyperparameter configurations as well as data augmentations. For the brief description for each model, please refer to Table \ref{tab:accall}. The black dot represents the median, and the error bars show the 16--84 percentile range.}
\label{fig:accall}
\end{figure*}

\begin{deluxetable*}{llccc}
\tablecaption{Performance of models improved with data augmentation and hyperparameter tuning}
\label{tab:accall}
\tablewidth{0pt}
\tablehead{
\colhead{Name} & Description &\colhead{Accuracy} & \colhead{F1-score} & \colhead{AUC}
}
\startdata
\texttt{Fiducial3}\tablenotemark{a} & Mild data augmentation (factor of 3)  & $0.538^{+0.00}_{-0.00}$  & $0.350^{+0.08}_{-0.08}$ & $0.517^{+0.07}_{-0.08}$\\
\texttt{Fiducial28} & Aggressive augmentation (factor of 28) &$0.597^{+0.03}_{-0.05}$ & $0.582^{+0.04}_{-0.15}$& $0.633^{+0.03}_{-0.05}$\\
\texttt{LReLU} & LeakyReLU & $0.610^{+0.03}_{-0.04}$ & $0.600^{+0.03}_{-0.09}$  & $0.64^ {+0.02}_{-0.03}$ \\
\texttt{AdamBatch64} & Batch size=64& $0.589^{+0.03}_{-0.05}$ & $0.568^{+0.05}_{-0.17}$ & $0.623^{+0.03}_{-0.06}$  \\
\texttt{AdamBatch256} & Batch size=256 & $0.579^{+0.05}_{-0.04}$ & $0.532^{+0.09}_{-0.18}$ & $0.635^{+0.02}_{-0.06}$ \\
\texttt{Adam0.01} & Learning rate=0.01 & $0.537^{+0.02}_{-0.00}$ & $0.349^{+0.17}_{-0.00}$ & $0.512^{+0.07}_{-0.01}$   \\
\texttt{Adam0.0005} & Learning rate=0.0005 & $0.618^{+0.02}_{-0.04}$ & $0.611^{+0.02}_{-0.08}$ & $0.654^{+0.02}_{-0.02}$  \\
\texttt{RAdamBatch64} & RAdam; Batch size=64 & $0.605^{+0.02}_{-0.03}$ & $0.595^{+0.02}_{-0.07}$ & $0.638^{+0.02}_{-0.02}$ \\
\texttt{RAdamBatch128} & RAdam & $0.618^{+0.01}_{-0.02}$ & $0.612^{+0.02}_{-0.03}$ & $0.649^{+0.02}_{-0.02}$ \\
\texttt{RAdamBatch256} & RAdam; Batch size=256 & $0.632^{+0.02}_{-0.02}$ & $0.629^{+0.02}_{-0.02}$ & $0.665^{+0.02}_{-0.02}$ \\
\texttt{DilationRate5} & Dilation rate=5 pixels in the first Conv2D layer & $0.642^{+0.02}_{-0.02}$ & $0.640^{+0.02}_{-0.02}$ & $0.664^{+0.02}_{-0.02}$ \\
\texttt{DilationRate10} & Dilation rate=10 pixels & $0.648^{+0.02}_{-0.02}$ & $0.645^{+0.02}_{-0.02}$ & $0.668^{+0.02}_{-0.02}$ \\
\texttt{DilationRate15} & Dilation rate=15 pixels & $0.640^{+0.01}_{-0.01}$ & $0.637^{+0.01}_{-0.02}$ & $0.660^{+0.01}_{-0.02}$ \\
\texttt{ConvLayer4} & four convolution layers & $0.559^{+0.05}_{-0.02}$ & $0.472^{+0.13}_{-0.12}$ & $0.598^{+0.05}_{-0.07}$ \\
\texttt{TrainRatio0.4} & Train:Validation:Test=48\%:12\%:40\% & $0.602^{+0.02}_{-0.05}$ & $0.589^{+0.03}_{-0.14}$ & $0.637^{+0.02}_{-0.03}$ \\
\texttt{CombineAll} &  LeakyReLU; RAdam; Batch Size=256 & $0.649^{+0.02}_{-0.02}$ & $0.646^{+0.02}_{-0.02}$ & $0.668^{+0.02}_{-0.02}$ \\
\enddata
\tablenotetext{a}{For the fiducial configuration, the hyperparameter setting is as follow: ReLU, Adam, batch size=128, learning rate=0.001, three convolution layers with no dilation, and the splitting ratios of training, validation, and test sets of 64\%, 16\%, and 20\%.}
\tablecomments{The 16--84 percentile range is provided as the uncertainty of the evaluation metrics.}
\end{deluxetable*}

\begin{figure*}[ht!]
\centering
\includegraphics[width=\linewidth]{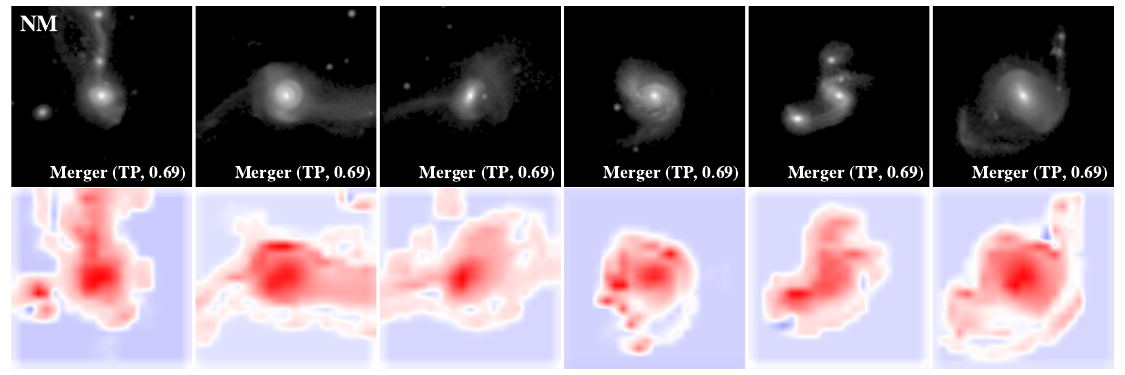}
\caption{Examples of True Positive (TP) cases from \texttt{CombineAll}, where TP refers to mergers correctly identified as such. Merger images and their corresponding Grad-CAM images are presented in the top and bottom rows, respectively. The Grad-CAM images are normalized to a range of -1 and 1 and are color-mapped using a blue-white-red color scheme, with redder indicating stronger model attention.
The model highlights both faint tidal features and bright cores, suggesting that faint tidal features contribute to merger identification.}
\label{fig:GCAM}
\end{figure*}

As described in Section \ref{sec:method:hyperparams}, we investigated various hyperparameter settings beyond the fiducial one, and examined their impact on model performance to identify the optimal combination of hyperparameters.
These alternative settings were applied to the dataset augmented by a factor of 28.
As summarized in Figure \ref{fig:accall} and Table \ref{tab:accall}, most alternative settings outperformed the fiducial one, yielding higher accuracy and reduced variability.
Although not shown, the high failure rates observed in \texttt{Fiducial3} and \texttt{Fiducial28} were significantly reduced with the improved models,
particularly in \texttt{DilationRate10}, where the fractions of failed training, overfitting, and successful training were 1.2\%, 0.1\%, and 98.7\%, respectively.
\texttt{RAamBatch256} also showed a low fraction for failed training, though the overfitting fraction remained high.
The models with a dilated convolution generally displayed more stable training histories, while others, as well as the fiducial ones, exhibited increases in validation loss, suggesting potential overfitting.
Not surprisingly, the best performance among the improved models was achieved by \texttt{DilationRate10}, with the median accuracy or F1-score of 65\% and the median AUC approaching 67\%.
The substantial improvement with a dilated convolution suggests that merger-induced features are more effectively captured with an expanded receptive field of an optimal size.
Further exploration of multiple dilation rates, rather than a single dilation rate, could provide additional benefits by capturing features across different scales.

By combining the hyperparameter settings that show better performance than the fiducial one, we determined the optimized model, referred to as \texttt{CombineAll}, which combines \texttt{LRelu}, \texttt{RAdamBatch256}, and \texttt{DilationRate10}.
The full configuration is detailed in Table \ref{tab:best}.
The performance of the \texttt{CombineAll} model was measured with an accuracy of 65\%, an F1-score of 65\%, and an AUC of 67\%.
The \texttt{CombineAll} model likely benefits from employing LReLu and Adam alongside a batch size of 256, enhancing training stability and convergence, which is anticipated outcome given the description of each hyperparameter in Section \ref{sec:method:hyperparams}.
Additionally, the significant performance improvement from incorporating dilated convolution suggests that the distinguishing features for classifying mergers and non-mergers are derived predominantly from global patterns rather than localized details, which may include tidal features.

While the overall performance of the \texttt{CombineAll} model is not significantly better than that of the \texttt{DilationRate10} model, the fraction of failed training in \texttt{CombineAll} was much lower (i.e., 1/1000 compared to 13/1000 for \texttt{DilationRate10}), suggesting that \texttt{CombineAll} is more robust across a wider range of datasets than \texttt{DilationRate10}.
While this level of performance of the \texttt{CombineAll} model is not high enough to be considered satisfactory, it is sufficient to demonstrate the feasibility of building a CNN model for classifying galaxy mergers from deep galaxy images.
One reason the model cannot achieve higher accuracy is due to incomplete labeling.
In some cases, mergers are mislabeled as non-mergers because the merger tree fails to identify them.
The issue arises when the halo finder fails detecting halos, particularly during the merging process.

Using the optimized configuration, we analyzed Grad-CAMs to identify which features were key in distinguishing mergers from non-mergers.
Grad-CAMs are generated using the output from the last convolutional layer after batch normalization is applied, producing a 2D feature map right before it is passed to the fully connected layer.
Figure \ref{fig:GCAM} displays the Grad-CAMs of six example galaxies from the true positive group (mergers correctly predicted by the model).
While the bright cores in galaxies are predominantly highlighted, indicating their influence on the model's decision, faint tidal features also appear to play a role.
To further investigate this, we conducted additional experiments, training the optimized model (\texttt{CombineAll}) on images processed to emphasize different regions (see Section \ref{sec:SBM} and Figure \ref{fig:SBM}) in the following section.

\begin{figure}[ht!]
\centering
\includegraphics[width=\linewidth]{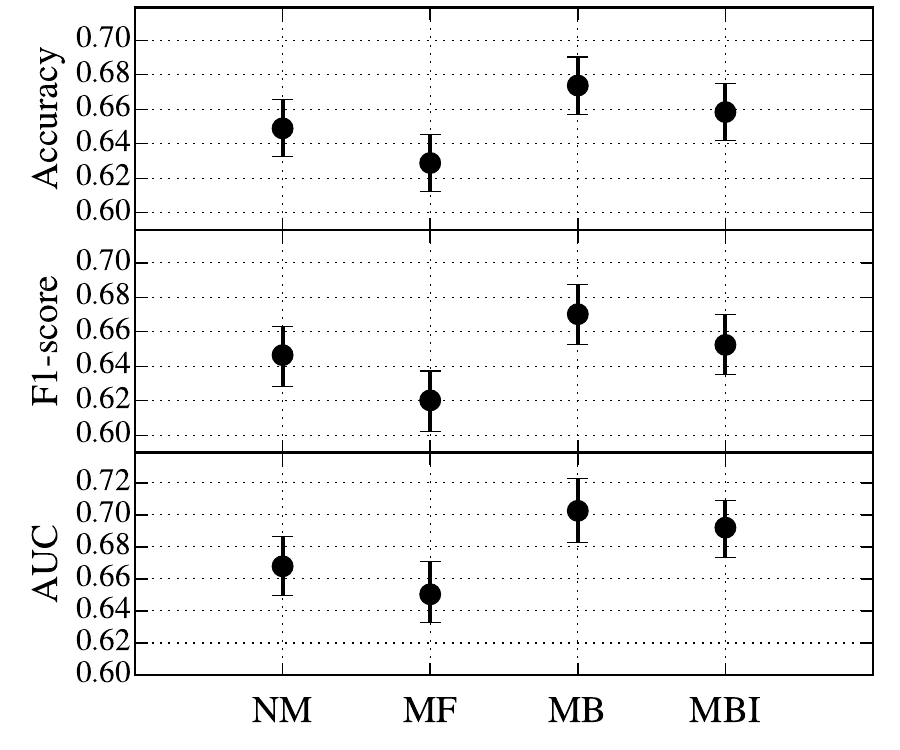}
\caption{Comparison of the performance between the NM, MF, MB, and MBI models. The MB model exhibits the highest performance, suggesting that low-surface brightness features are closely related to merger properties.}
\label{fig:acc4}
\end{figure}

\begin{deluxetable}{lccc}
\tablecaption{The performance of the optimized model (\texttt{CombineAll}\tablenotemark{b}) trained on four datasets with different processings methods. NM represent the original images, MF denotes images with faint features masked ($\geq26{\rm mag/arcsec^2}$), MB denotes images with bright features masked ($<26{\rm mag/arcsec^2}$), and MBI represents the inverted version of MB.}
\label{tab:acc4}
\tablewidth{0pt}
\tablehead{
\colhead{Name} &\colhead{Accuracy} & \colhead{F1-score} & \colhead{AUC}
}
\startdata
NM & $0.649^{+0.017}_{-0.017}$ & $0.646^{+0.017}_{-0.018}$ & $0.668^{+0.019}_{-0.018}$ \\
MF & $0.629^{+0.017}_{-0.017}$ & $0.620^{+0.017}_{-0.018}$ & $0.650^{+0.020}_{-0.018}$ \\
MB & $0.674^{+0.017}_{-0.017}$ & $0.670^{+0.017}_{-0.018}$ & $0.703^{+0.020}_{-0.020}$ \\
MBI & $0.658^{+0.017}_{-0.017}$ & $0.652^{+0.018}_{-0.017}$ & $0.692^{+0.017}_{-0.019}$ \\
\enddata
\tablenotetext{b}{For the \texttt{CombineAll} configuration, the hyperparameter setting is as follow: LReLU, RAdam, batch size=256, learning rate=0.001, three convolution layers with the first convolution layer having a dilation rate of 10, and the splitting ratios of training, validation, and test sets of 64\%, 16\%, and 20\%.}
\tablecomments{The 16--84 percentile range is provided as the uncertainty of the evaluation metrics.}
\end{deluxetable}

\begin{figure*}[ht!]
\centering
\includegraphics[width=\linewidth]{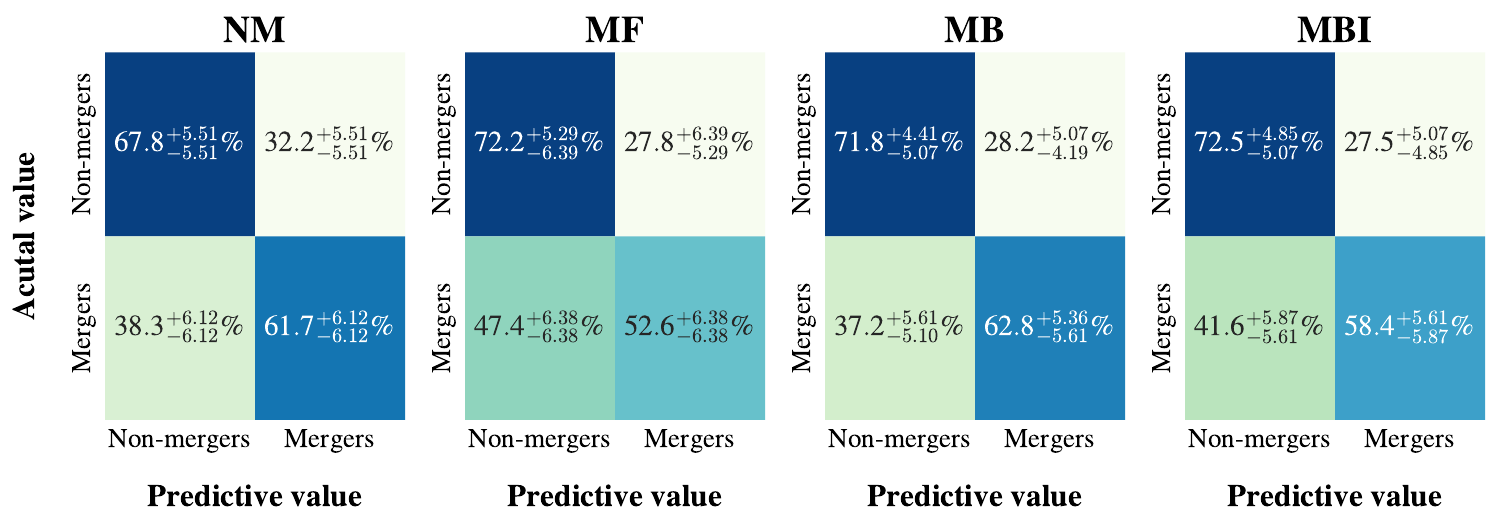}
\caption{Confusion matrices of the optimized model (\texttt{CombineAll}) trained on four datasets of NM (No Masking), MF (Masking of Faint features), MB (Masking of Bright features), and MBI (Masking of Bright features and Inverted).}
\label{fig:Confusion4}
\end{figure*}

\subsection{Model optimization: data processing}\label{sec:res:SBM}

Four different sets of galaxy image are prepared: original images with No Masking (NM), those with Masking of Faint features (MF), those with Masking of Bright features (MB), and those with Masking of Bright features and Inverted (MBI).
The division between faint and bright features is set at $26\ {\rm mag\,arcsec^{-2}}$.
The example images are shown in Figure \ref{fig:SBM} in Section \ref{sec:SBM}.
The performances of the models trained on these four datasets are presented in Figure \ref{fig:acc4} and Table \ref{tab:acc4}.

The model achieves its highest performance (an AUC of 70\%) when trained on the MB dataset.
This is fairly good performance, especially considering the complexity of determining whether a galaxy has experienced mergers based on galaxy images.
The MF model showed its lowest performance, with a 2--3\% decrease across all evaluation metrics compared to the NM model, and a drop of up to $\sim$5\% when compared to the MB model.
It is interesting that the model performed better when trained on the dataset with bright features masked (MB) compared to when trained on the dataset with no masking (NM).
These findings support the idea that faint tidal features may play a significant, perhaps even crucial role in identifying mergers.
When trained on the dataset with faint features further emphasized through inversion (MBI), the model's performance slightly decreased.
While masking bright features to emphasize faint ones seems to have been beneficial, the additional step of inversion may have introduced some complications.
One possibility is that faint features produced by ``non-mergers,'' such as mergers with a mass ratio below 1/10 or flybys, are unnecessarily emphasized, thereby amplifying confusion in the MBI model.

The comparison of confusion matrices of the four models, averaged over 1000 instances of each, in Figure \ref{fig:Confusion4} shows that the lower performance of the MF model (compared to the NM model) is primarily due to misclassifying mergers as non-mergers (47.4\% vs. 38.3\%).
This reinforces the idea that mergers are more accurately identified when faint features are present.
Conversely, the MF model's better identification of non-mergers (72.2\% compared to 67.8\% of the NM model) suggests that faint features complicate distinguishing non-mergers.
This naturally leads to an expectation that the MB and MBI models would exhibit similar accuracy for non-mergers as the NM model, i.e., a lower accuracy than the MF model.
However, both the MB and MBI models achieve non-merger identification rates of 71.8\% and 72.5\%, respectively, which are higher than the NM model and comparable to the MB model.
One possible explanation for the MB and MBI models' more accurate non-merger identification compared to the NM model is that key clues for identifying non-mergers appear more clearly when bright features are masked.

\begin{figure*}[ht!]
\centering
\includegraphics[width=0.8\linewidth]{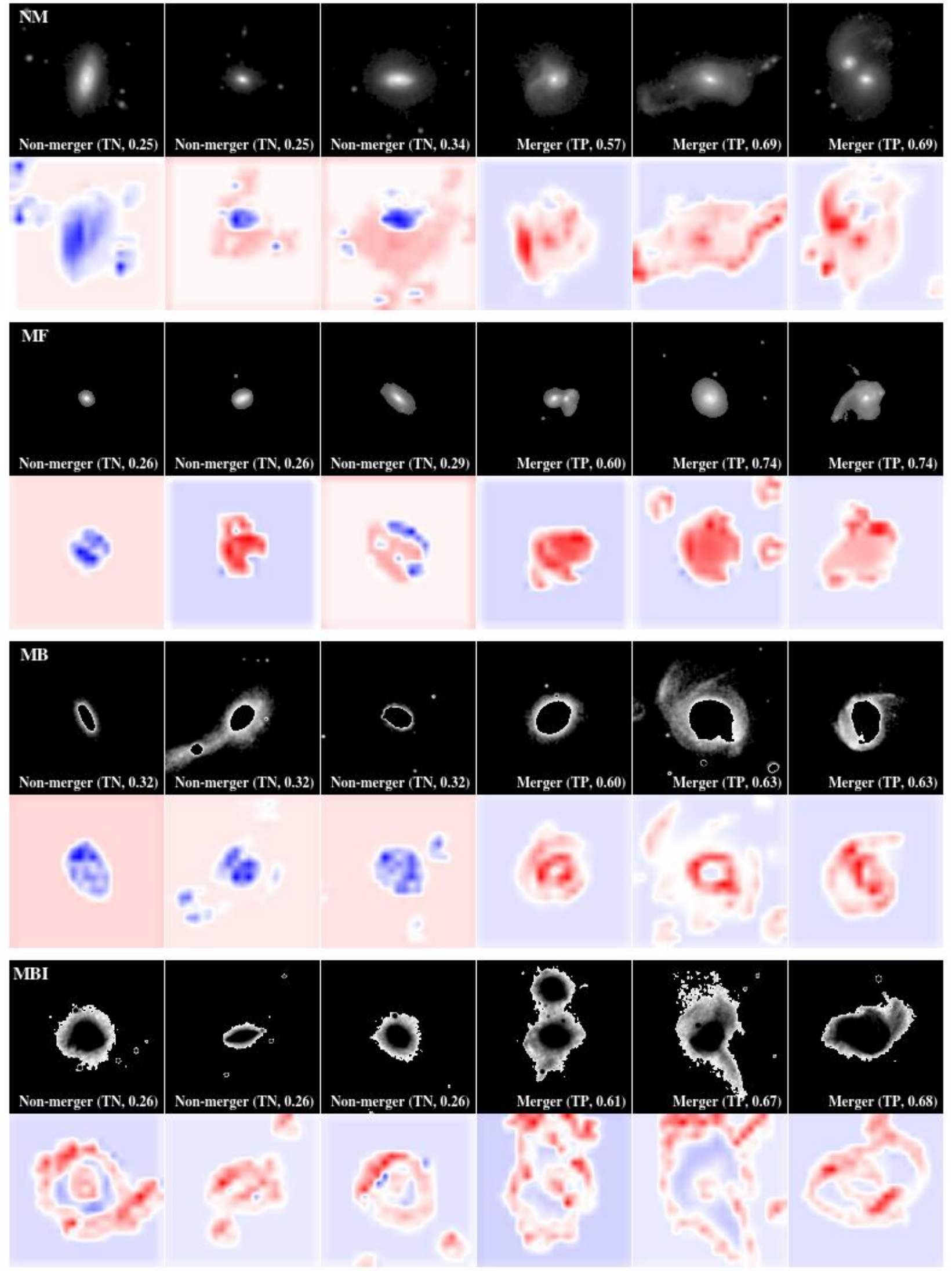}
\caption{Representative Grad-CAM images of the four models, the NM, MF, MB, and MBI models from top to bottom, with corresponding input galaxy images above.
The true label for each case is provided along the model prediction in parenthesis (TN and TP denote true negative and true positive, respectively; the number represents the model output, closer to 0 for non-mergers and closer to 1 for mergers). It is in the order of the output.}
\label{fig:gradcam_all}
\end{figure*}

\begin{figure*}[ht!]
\centering
\includegraphics[width=0.8\linewidth]{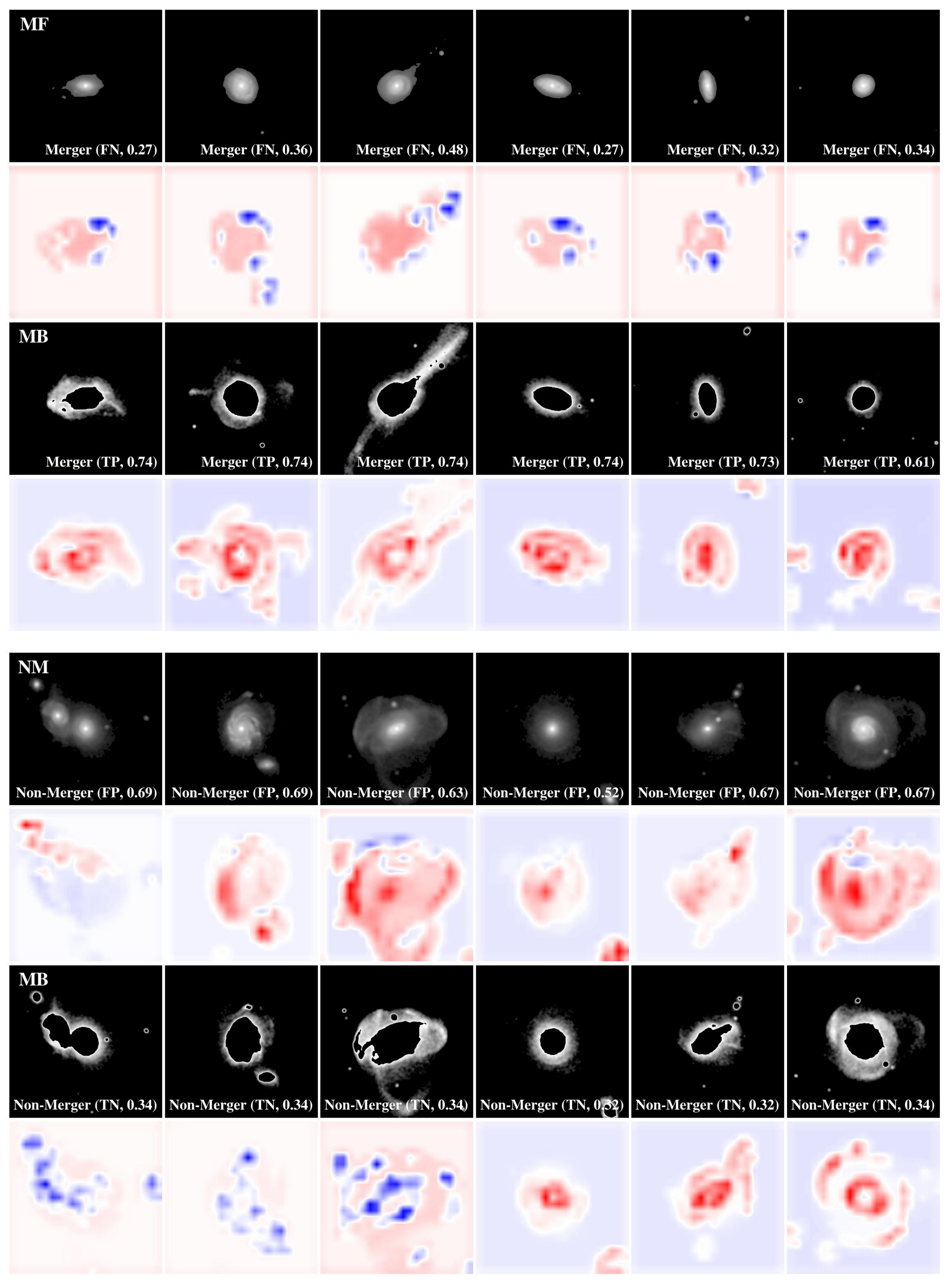}
\caption{Top four rows show merger examples that were successfully classified by the MB model but not in the MF model, whereas the bottom four rows display non-mergers examples that were misclassified by the NM model but correctly classified by the MB model.
The true label and merger prediction with the model output are provided in the bottom of each galaxy cutout image, as in Figure \ref{fig:gradcam_all}.}
\label{fig:Missclassification}
\end{figure*}

To better understand each model's performance, we examine Grad-CAM images across the four models.
Figure \ref{fig:gradcam_all} presents representative examples, with non-mergers and mergers arranged from left to right.
In the NM model (top two rows), mergers are identified based on both the central region of the galaxy and the overall structure, including tidal features, as already shown in Figure \ref{fig:GCAM}.
In some cases, the model prioritizes tidal features over the bright central region.
For non-mergers, the model tends to disregard the bright central region.
It is also observed that the Grad-CAM images of non-mergers appear with a red background.
This suggests that the model interprets a relatively clean background, free from neighboring galaxies and tidal features, as evidence of a non-mergers.
Consequently, the main limitation would arise from misclassifications of mergers whose morphological disturbances have faded and non-mergers retaining residual features from interactions that are not identified as mergers based on our merger definition.

The following two rows in Figure \ref{fig:gradcam_all} show the case of the MF model, where a significant portion (fainter than $26\,\mathrm{mag\,arcsec^{-2}}$) is obscured.
As a result, mergers and non-mergers appear quite similar in the images.
Upon closer examination, merger typically exhibit more extended, asymmetric shapes, along with a higher incidence of multiple cores and neighboring galaxies.
The model seems to leverage those morphological distinctions for classification.
The Grad-CAM images of non-mergers often display a red background, as seen in the NM model case.

In the MB model (fourth and third panels from the bottom), mergers tend to be identified based on the remaining unmasked galaxy structure along with information from the mask boundary, whereas non-mergers are classified with greater reliance on the background compared to the previous two models.
Since our study uses images that do not include foreground and background galaxies, the MB model's heavy reliance on the background region suggests that its performance may degrade when applied to more realistic images containing such galaxies, as indicated by \citet{Bottrell2019} and \citet{Bottrell2022}.
This underscores the need for further investigation into the impact of foreground and background galaxies on classification performance, which we plan to explore in future work.

The MBI model is trained on images where tidal features are well visible both from mergers and non-mergers (second row from the bottom), and their Grad-CAM images (bottom row) show comparable patterns.
It is notable that, unlike other models, the model tends to focus on tidal features rather than background regions when identifying non-mergers.
This can be attributed to the brightness inversion, with which faint features emerge more clearly, allowing for a more nuanced interpretation and utilization of them.
Since the MBI model primarily focuses on tidal features rather than background, it may be less affected by the presence of foreground and background galaxies compared to other models.

To further understand the MB model's success, we examine cases correctly identified by the MB model but misclassified by the other models. 
Given that the MB model clearly outperforms the MF model in identifying mergers and the NM model in identifying non-mergers, we compare the model pairs MB-MF and MB-NM.
The top four rows in Figure \ref{fig:Missclassification} display mergers misclassified by the MF model but correctly classified by the MB model.
Many of these cases exhibit prominent faint features but appear featureless when their faint features are obscured, which reduces the performance of the MF model.
Surprisingly, with bright features masked, the MB model appears to be able to effectively capture faint features even in cases where faint features are not particularly prominent (the three cases on the right).
Similarly, the bottom four rows in Figure \ref{fig:Missclassification} present non-merger examples that the NM model misclassifies while the MB model correctly identifies.
Some of these cases exhibit prominent faint features, making them susceptible to being misclassified as mergers.
While both models focus on tidal features, their predictions differ, with the MB model making correct prediction.
This suggests that the MB model advances the interpretation of tidal features, likely facilitated by the emphasis on faint features through the masking of bright regions.
One additional effect of masking bright features is the occlusion of nearby satellites, which in some cases helps improve non-merger classification.

Although the highest-performing instance in each model yields comparable accuracy (i.e.,  $\sim70\%$ for accuracy), the MB model stands out with the highest median performance and the lowest variation across instances.
Its superior accuracy in identifying both mergers and non-mergers strongly supports the idea that faint tidal features contribute meaningfully to merger classification.
Moreover, this highlights that the method of image processing can significantly influence model performance.

\subsection{Additional tests}\label{sec:additionaltests}

We conducted additional tests to examine the impact of photometric noise, filter dependence, and sample size.
To save space, we opt to exclude figures and provide only a brief summary of the results.
We trained a model on images with bright regions masked (as in the MB model), but with random noise added at a level comparable to LSST, with a 3$\sigma$ surface brightness limit of $\sim29\,\mathrm{mag\,arcsec^{-2}}$ averaged over the LSST filter set.
Another model was trained on MB-like images generated specifically in the LSST $r$-band.
Finally, we trained an additional model using an expanded sample, where the halo mass range was extended from $11.9<\log_{10} M_{\rm halo}/M_\odot < 12.3$ to $11.2<\log_{10} M_{\rm halo}/M_\odot < 12.3$.
This increased the number of galaxies from 151 to 1008 and the number of mergers from 70 to 328.
We applied the same data augmentation and image processing in the $r$-band for this model.

The changes in performance across these models, relative to the fiducial MB model, are marginal, with median F1-score values of 0.67, 0.66, and 0.66 (c.f., 0.67 for the fiducial MB model).
This suggests that photometric noise has minimal impact on performance, although the implemented noise may underestimate real observational conditions.
The model trained on $r$-band images performs slightly worse, likely because tidal features are less prominent in the $r$-band than in the $K$-band. 
While the difference is not significant, this underscores the importance of training models on images in the target observational band.
Contrary to expectations, increasing the sample size does not lead to notable performance gains, though it does reduce performance variability across model instances.
This implies that further improvement may hinge more on advancements in model architecture--such as adopting ResNet \citep{He2015}, GANs \citep{Goodfellow2014}, or attention mechanisms \citep{Vaswani2017}--than on simply expanding the dataset.

We also tested how the model performance will change at a higher redshift.
We trained the four models (NM, MF, MB, and MBI) at $z=0.42$, reprocessing the images to account for redshift dimming and the increased pixel scale (physical scale per pixel).
As expected, galaxies at a higher redshift appear smaller and their faint features tend to vanish below the detection limit.
This degradation in image quality would negatively impact model performance, particularly in identifying mergers.
Consistent with this expectation, all four models showed reduced performance at $z=0.42$.
Specifically, the confusion matrices reveal a $\sim5\%$ drop in merger identification rate for the MB and MBI models, likely due to the blurring critical merger features (i.e., tidal features).
This result is consistent with \citet{Bickley2024b}, which reported a decline in model performance with increasing redshift over the range of 0.036 to 0.256.
Although overall performance declines, the MB model remains the best-performing model at $z=0.42$.
However, given its relatively greater decline, there is likely a redshift threshold beyond which the NM model starts to outperform it.
At higher redshifts, where information loss increases, it may become more advantageous to utilize all available information without masking.

\begin{figure*}[ht!]
\centering
\includegraphics[width=0.57\linewidth]{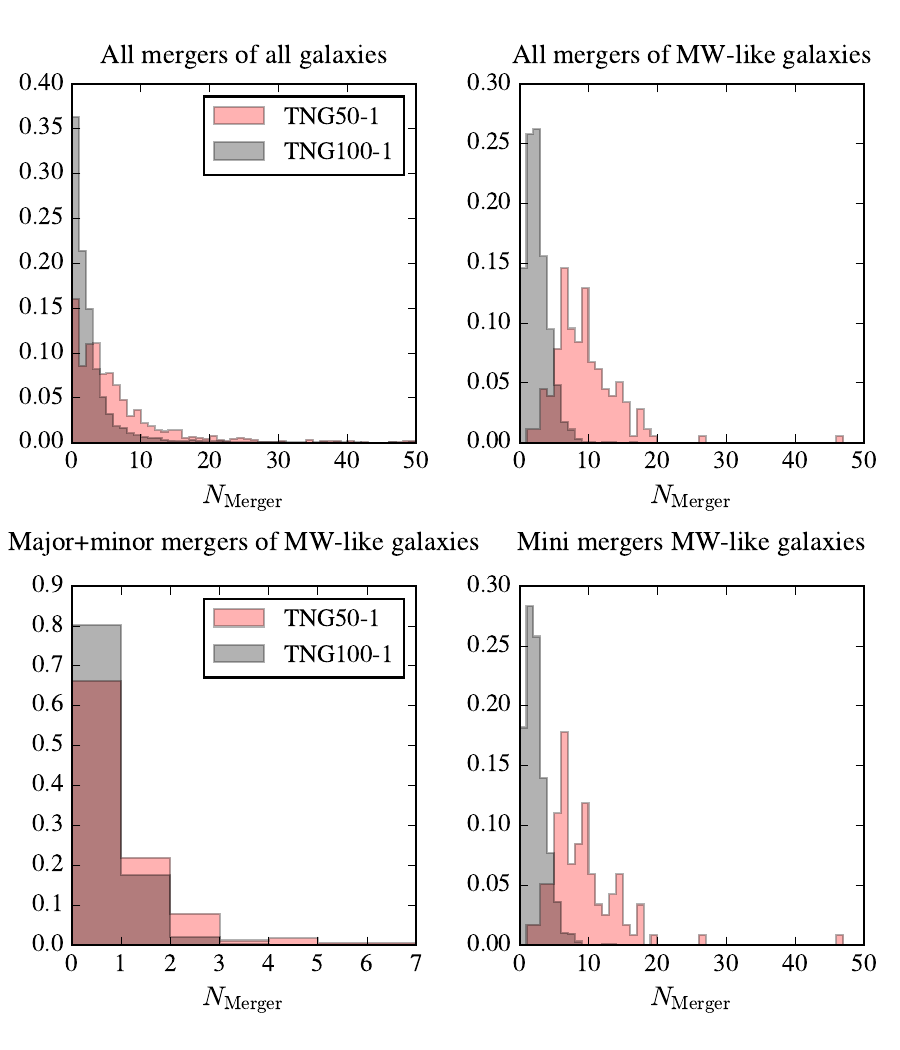}
\caption{The top row shows normalized histograms of the total number of mergers for all galaxies (left) and MW-like galaxies (right) in TNG50 (red) and TNG100 (black). The bottom row displays normalized histograms of the number of major and minor mergers (stellar mass ratio $\mu\ge0.1$) in the left panel and “mini” mergers ($\mu<0.1$) in the right panel for MW-like galaxies in TNG50 (red) and TNG100 (black).}
\label{fig:Hist_Nm}
\end{figure*}

\begin{figure*}[ht!]
\centering
\includegraphics[width=0.9\linewidth]{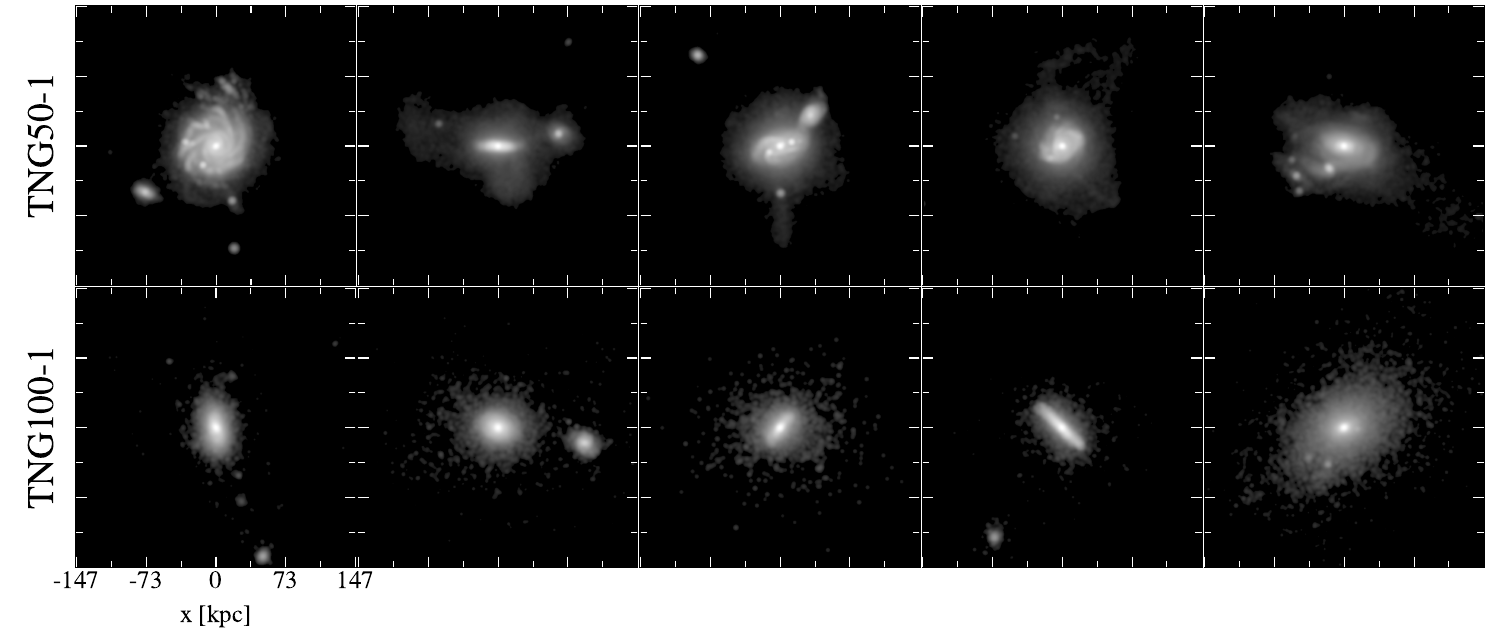}
\caption{Non-mergers (those experiencing neither major nor minor mergers) in TNG50 (top) and TNG100 (bottom), with similar stellar masses ($10.76<\log_{10} M_{\rm st}/M_{\odot}<10.91$). Non-mergers in TNG50 appear more disturbed than those in TNG100, likely due to the presence of “mini” mergers (stellar mass ratio $\mu<0.1$), which are less resolved in TNG100 due to its lower resolution.}
\label{fig:tng50_tng100}
\end{figure*}

\subsection{Comparison with previous studies}

As mentioned in Section \ref{sec:intro}, there have been a few studies that utilized TNG100 to develop merger-classifying ML models for various surveys including LSST \citep[e.g.,][]{Wang2020,Bickley2021,Ferreira2022,Bottrell2022,Ferreira2024}.
The models developed in these studies achieve high classification accuracies of 84--88\%, significantly surpassing our results.
This discrepancy cannot be fully explained by differences in sample selection, merger definitions, or image quality alone.
Instead, it may be attributed to variations in the physical processes realized at different resolutions.
Specifically, mergers and fly-bys--particularly those involving low-mass galaxies--are less frequently resolved in TNG100 than in TNG50.
This likely simplifies merger classification in TNG100 compared to TNG50.

To investigate this further, we compare the number of mergers in TNG100 and TNG50.
Figure \ref{fig:Hist_Nm} shows the distribution of merger events experienced by galaxies, revealing that galaxies in TNG50 undergo a greater number of mergers (top left panel).
When restricting the sample to MW-like galaxies--the primary focus of our study and others--the discrepancy between TNG100 and TNG50 becomes even more pronounced (top right panel).
This difference primarily arises from the presence of ``mini'" mergers (stellar mass ratio $M_1/M_2=\mu<1/10$), as shown in the bottom row of Figure \ref{fig:Hist_Nm}.
The bottom left panel of Figure \ref{fig:Hist_Nm} compares the frequency of major and minor mergers ($\mu\ge0.1$), while the right panel focuses on mini mergers.
Because mini mergers are often excluded from merger counts (as in ours and other studies), galaxies that have only experienced mini mergers may introduce ambiguity in classification, particularly if these mergers generate detectable tidal features.
The bottom right panel of Figure \ref{fig:Hist_Nm} demonstrates that galaxies that experience mini mergers only are more prevalent in TNG50 with a greater number of such mergers.
To further examine this effect, we compare images of non-mergers galaxies with similar stellar masses across the two simulations.
As illustrated in Figure \ref{fig:tng50_tng100}, the higher resolution of TNG50 results in more pronounced tidal features due to mini mergers, increasing the complexity of classification.

In summary, tidal features are underrepresented in TNG100, which likely explains the superior classification performance observed in previous studies.
This interpretation is further supported by \citet{Omori2023}, who utilized TNG50 to classify mergers in Subaru HSC-SSP data (with a surface brightness limit of $28.5\,\mathrm{mag\,arcsec^{-2}}$ in the $g$-band) and reported an accuracy of $\sim76\%$, lower than the works based on TNG100.
Their higher performance, compared to ours, likely reflects their stricter selection criteria for mergers and non-mergers, defined by narrower time windows--within 0.5 Gyr of the closest merger event for mergers and beyond 3 Gyr for non-mergers--allowing a clearer distinction between the two classes.

These considerations highlight the importance of high-resolution simulations, such as TNG50, when developing merger classification models for LSST-like deep imaging, where faint, low-surface-brightness tidal features are observable.
Furthermore, it would be worthwhile to broaden the definition of mergers to include mini mergers, as subtle tidal signatures can be produced even by such minor interactions.
However, the Rodriguez merger catalog used in this study classifies mergers only for mass ratios of $>1/4$, $>1/10$, and any other mass ratio, without a detailed distinction for mini mergers ($<1/10$).
Including all mass ratios would result in every galaxy being labeled as a merger, leaving no non-merger counterparts for comparison.
A more comprehensive analysis that treats mini mergers as a distinct class would require reconstructing the merger catalog with finer mass-ratio bins, which is beyond the scope of the present work.
We plan to address this issue in future studies, particularly in re-defining mergers responsible for the tidal features detectable in LSST images.

\section{Summary and outlook}\label{sec:summary}

In this study, we demonstrated the feasibility of developing a simple CNN model that identifies mergers in LSST-like deep images of low-redshift ($z=0.2$) galaxies using the TNG50 simulation.
To simplify the problem, we focused on 151 Milky Way-like central galaxies in field environments, which are expected to have relatively simple merger histories.
We utilized rest-frame $K$-band images that closely trace tidal features.
The model was optimized through data augmentation and hyperparameter tunning with a dilated convolution layer significantly enhancing the model performance.
The Grad-CAM method reveals that the optimized model (CombineAll) identifies mergers by leveraging faint tidal features.
Notably, the optimized model achieves the best performance when trained on images with bright features (< 26 ${\rm mag\,arcsec^{-2}}$) masked (the MB model), suggesting that faint tidal features serve as effective discriminators between merger and non-mergers.

To further improve the model, we plan to increase the sample size by extending the mass range and incorporating satellite galaxies.
Additionally, comparing different environments is one of the future directions for this research.
While we used a simple CNN model composed of three convolutional layers, we can employ more complex architectures, such as multispectral or multi-channel CNNs for multi-band input images, as well as the Inception module to utilize various dilated convolution rates.
This approach may lead to the development of a more advanced CNN model capable of providing detailed information about mergers (e.g., mass ratio, time since closest encounter, etc.).
A more comprehensive analysis would also involve treating mini mergers as a distinct class, which requires reconstructing the merger catalog with finer mass-ratio bins; we plan to address this issue in future studies, particularly by re-defining mergers in a way that accounts for the tidal features detectable in LSST images.
Consequently, it would broaden the research scope of galaxy formation and evolution by facilitating in-depth studies of mergers based on observational data, which has primarily relied on simulation data to date.

\begin{acknowledgments}
We are grateful to the anonymous referee and the editor for comments that have improved this paper.
This work was supported by the National Research Foundation of Korea (NRF) grant funded by the Korea government (MSIT) (No. 2022M3K3A1093827, 2022R1A4A3031306).
J.L. is supported by the National Research Foundation of Korea (NRF-2021R1C1C2011626).
\end{acknowledgments}

\vspace{5mm}
\facilities{GPU computing resources in the department in the Department of Astronomy and Atmospheric Sciences, College of Natural Sciences, Kyungpook National University}
\software{HEALPix \citep{healpix1,healpix2}, Tensorflow \citep{tensorflow2015-whitepaper}, Keras \citep{chollet2015keras}}, Astropy \citep{astropy:2018} 

\bibliography{name}

\FloatBarrier

\appendix 
\section{Model Architecture}\label{app:model}
The model architectures for the fiducial model (adopted for \texttt{Fiducial3} and \texttt{Fiducial28}) and the optimized model (\texttt{CombineAll}) are summarized in Tables \ref{tab:Fiducial} and \ref{tab:best}.
The model architectures for other models presented in Figure \ref{fig:accall} and Table \ref{tab:accall} are provided at \url{https://github.com/yeonkyung-lab/Merger-classifying-CNN-model-for-LSST}.

\begin{deluxetable*}{llllll}
\tablecaption{Architecture of \texttt{Fiducial} model}
\tablewidth{0pt}
\label{tab:Fiducial}
\tablehead{
\colhead{Layers} & \colhead{Properties} & \colhead{Stride} & \colhead{Padding} &\colhead{Output Shape} & \colhead{Parameters}
}
\startdata
Input  & $1\times 600\times 600$ & -  & (1,600,600) & (1,600,600) &0 \\
\toprule[2pt]
Convolution(2D) & \makecell[l]{Filters : 8\\ Kernel : $15\times 15$ \\ Activation : ReLU} & $8\times8$  & Valid & (8,74,74) & 1803 \\
\hline
Batch Normalization & - & -   & -  & (8,74,74) & 32 \\ 
\hline
MaxPooling(2D) & Kernel : $2\times2$  & $2\times2$ & Valid  & (8,37,37) & 0  \\ 
\hline
Dropout & Rate : 0.5  & -& -  & (8,37,37)& 0 \\ 
\toprule[2pt]
Convolution(2D) & \makecell[l]{Filters : 16 \\ Kernel : $3\times3$\\ 
Activation : ReLU} & $1\times1$  & Same & (16,37,37) & 1168 \\ 
\hline
Batch Normalization & - & -   & -  & (16,37,37) & 64 \\ 
\hline
MaxPooling(2D) & Kernel : $2\times2$  & $2\times2$ & Valid  & (16,18,18) & 0  \\ 
\hline
Dropout & Rate : 0.5  & -& -  & (16,18,18)& 0 \\ 
\toprule[2pt]
Convolution(2D) & \makecell[l]{Filters : 32 \\ Kernel : $3\times3$\\ Activation : ReLU} & $1\times1$  & Same & (32,18,18) & 4640 \\ 
\hline
Batch Normalization & - & -   & -  & (32,18,18) & 128 \\ 
\hline
MaxPooling(2D) & Kernel : $2\times2$  & $2\times2$ & Valid  & (32,9,9) & 0  \\ 
\hline
Dropout & Rate : 0.5  & -& -  & (32,9,9)& 0 \\ 
\toprule[2pt]
Flatten & - & - & - & (2592) & - \\ 
\hline
Fully connected & \makecell[l]{Reg: L2(0.0001)\\Activation: Softmax } & - & - & (64) & 165952 \\ 
\hline
Fully connected & \makecell[l]{Reg: L2(0.0001)\\Activation: Softmax } & - & - & (32) & 2080 \\ 
\hline
Fully connected & Activation: Sigmoid & - & - & (1) & 33 \\ 
\toprule[0.5pt]
\enddata
\end{deluxetable*}

\begin{deluxetable*}{llllll}
\caption{The architecture of the optimized model (\texttt{CombineAll})}
\tablewidth{0pt}
\label{tab:best}
\tablehead{
\colhead{Layers} & \colhead{Properties} & \colhead{Stride} & \colhead{Padding} &\colhead{Output Shape} & \colhead{Parameters}
}
\startdata
Input  & $1\times 600\times 600$ & -  & (1,600,600) & (1,600,600) &0 \\ \hline
\toprule[1.5pt]
Convolution(2D) & \makecell[l]{Filters : 8 \\ Kernel : $7\times 7$\\ Activation :\\LeakyReLU($\alpha=0.01$) \\dilation rate=10} & $1\times1$  & Valid & (8,540,540) & 400 \\ \hline
Batch Normalization & - & -   & -  & (8,74,74) & 32 \\ \hline
MaxPooling(2D) & Kernel : $2\times2$  & $2\times2$ & Valid  & (8,270,270) & 0  \\ \hline
Dropout & Rate : 0.5  & -& -  & (8,270,270)& 0 \\ \hline
\toprule[2pt]
Convolution(2D) & \makecell[l]{Filters : 16 \\ Kernel : $5\times5$\\ Activation : \\LeakyReLU($\alpha=0.01$)} & $1\times1$  & Same & (16,90,90) & 3216 \\ \hline
Batch Normalization & - & -   & -  & (16,90,90) & 64 \\ \hline
MaxPooling(2D) & Kernel : $2\times2$  & $2\times2$ & Valid  & (16,45,45) & 0  \\ \hline
Dropout & Rate : 0.5  & -& -  & (16,45,45)& 0 \\ \hline
\toprule[2pt]
Convolution(2D) & \makecell[l]{Filters : 32 \\ Kernel : $3\times3$\\ Activation : \\LeakyReLU($\alpha=0.01$)} & $1\times1$  & Same & (32,23,23) & 4640\\ \hline
Batch Normalization & - & -   & -  & (32,23,23) & 128 \\ \hline
MaxPooling(2D) & Kernel : $2\times2$  & $2\times2$ & Valid  & (32,23,23) & 0  \\ \hline
Dropout & Rate : 0.5  & -& -  & (32,11,11)& 0 \\ \hline
\toprule[2pt]
Flatten & - & - & - & (3872) & - \\ \hline
Fully connected & \makecell[l]{Reg: L2(0.0001)\\Activation: Softmax } & - & - & (64) & 247872 \\ \hline
Fully connected & \makecell[l]{Reg: L2(0.0001)\\Activation: Softmax } & - & - & (32) & 2080 \\ \hline
Fully connected & Activation: Sigmoid & - & - & (1) & 33 \\ \hline
\toprule[0.5pt]
\enddata
\end{deluxetable*}

\end{document}